\documentstyle[psfig]{elsart}

\begin{document}
\begin{frontmatter}

\title{
Characterization of nuclear physics targets using Rutherford backscattering
and particle induced x-ray emission
}
\author
{Th. Rubehn\thanksref{EMAIL},}
\author
{ G.J. Wozniak,
 L. Phair,
 and L.G. Moretto
}
\address{
Nuclear Science Division\\
Ernest Orlando Lawrence Berkeley National Laboratory,\\
University of California, Berkeley, California 94720, USA
}
\author
 {Kin Man Yu}
\address{
Materials Sciences Division,\\
Ernest Orlando Lawrence Berkeley National Laboratory,\\
University of California, Berkeley, California 94720, USA
}
\thanks[EMAIL]{E-mail address: TRubehn@lbl.gov}
%
%
\date{\today}

\begin{abstract}
Rutherford backscattering and particle induced x-ray emission 
have been utilized to precisely characterize targets used in 
nuclear fission experiments. 
The method allows for a fast and non destructive determination
of target thickness, homogeneity and element composition.
\end{abstract}

\end{frontmatter}


\section{Introduction}
Many experiments in nuclear physics require targets with a precise
characterization. In particular, one needs a precise determination 
of quantities such as the target thickness, the homogeneity, and the 
amount and kind of impurities, in order to investigate 
rare processes or perform high accuracy measurements.

Recently, we have investigated $^3$He- and $^4$He-induced
nuclear fission of several compound nuclei at bombarding 
energies between 20 and 145 MeV measured at the 88-Inch 
Cyclotron of the Lawrence Berkeley National Laboratory 
\cite{Mor95,Rub96a,Rub96b,Rub96c}. 
To study the excitation energy dependence of the first chance 
fission probability, which is determined by subtracting similar 
cross sections of two neighboring isotopes \cite{Rub96c}, 
it is essential to 
measure the cumulative fission cross sections with high precision.
While statistical errors can be minimized by measuring a sufficiently 
large number of fission events, systematic errors, as for example
caused by uncertainties in the target thickness or uniformity, are of
particular concern and must be evaluated.

\begin{figure}[htb]
\centerline{\psfig{file=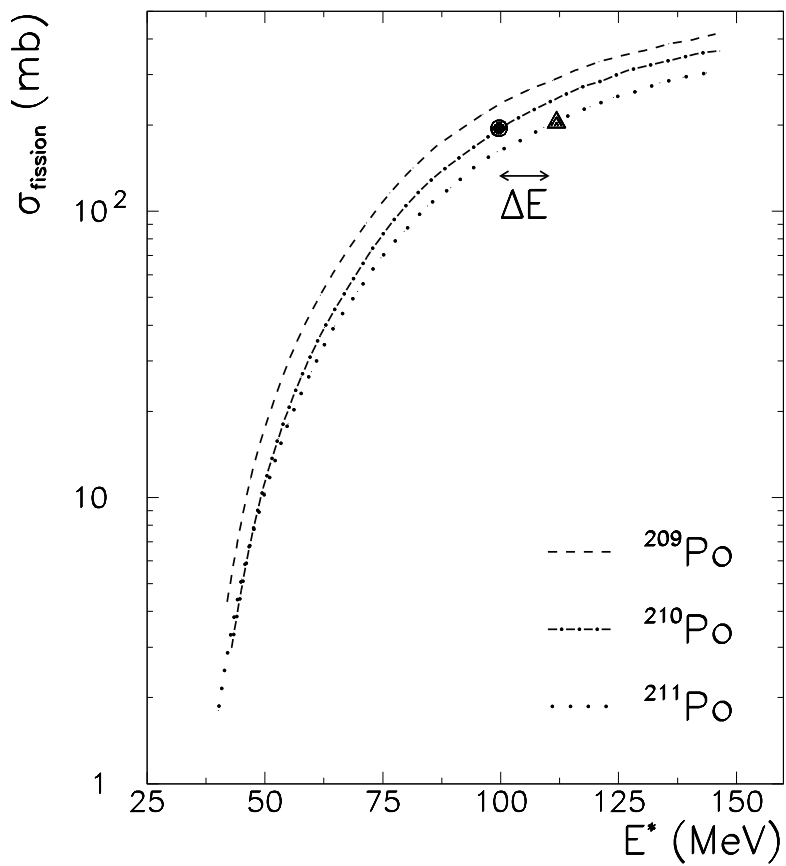,height=12cm}}
\caption{
 Schematic excitation functions for the cumulative fission of the 
 compound nuclei $^{209}$Po, $^{210}$Po, $^{211}$Po. The 
 first chance fission probability can be determined by 
 subtracting similar cross sections of the mother
 (triangle) and daughter nucleus (circle).
 }
\label{f1}
\end{figure}

To qualitatively illustrate the accuracy needed for our measurements,
we schematically show in Fig.~\ref{f1} the fission excitation 
functions of three neighboring lead isotopes \cite{Rub96c}. 
The first chance 
fission probability is determined by the difference in the cross
sections of the mother (triangle) and daughter nucleus (circle)
separated by the kinetic and the binding energies of
the evaporated neutron.
Due to the flattening of the curves 
at large excitation energies, the cross sections 
become more similar and thus precise cross sections
measurements are required.

The presence of contaminations from heavier elements 
represents another systematic
uncertainty in fission cross sections measured from targets made 
of lighter elements in the rare earth region: Due to their 
substantially lower fission barriers, even a small contamination from 
heavy elements ($<$ 1 ppm) can significantly increase the measured 
fission cross sections \cite{Rai67}. This effect is most
prominent at low excitation energies near the fission barrier of the 
lighter element.

In this paper, we report on the use of Rutherford backscattering 
and particle induced x-ray emission for a precise off-line 
characterization of targets used in nuclear physics experiments. 
Furthermore, we introduce a sensitive method to check the
relative accuracy of cross  section measurements.

\section{The Rutherford backscattering technique}

Rutherford scattering was studied at the beginning of the century 
by Rutherford \cite{Rut11}, Geiger and Marsden \cite{Gei13}.
Their experiments were purely of nuclear physics interest, i.e. 
they were designed to confirm the atomic model proposed by 
Rutherford. The analytical nature of the Rutherford backscattering 
method (RBS), however, 
was not fully realized until the late 1950s \cite{Rub57}.

For several decades, RBS has been used as a technique to 
characterize the surface and near surface properties 
of thin films of thicknesses between $\sim$100{\AA} 
and 1$\mu$m  (see e.g. Ref.~\cite{Wil78}). 
The major push to use this method has come from the need to analyze 
electronic materials like semiconductors \cite{Nav83,Wit83}. 
The technique has also been used to investigate ion
implantations into solids \cite{Wil84}.

As in the original experiments by Rutherford, Geiger and Marsden, 
the RBS technique analyzes the Coulomb interaction between a projectile 
of charge $Z_1 e$ and a target nucleus of charge $Z_{2} e$.
As we will briefly discuss in this section, the energy and scattering 
angle of the scattered particle provide information on the thickness, 
the nature of constituents, and the profile of the target.
A typical experimental setup requires a beam generating device 
(providing a collimated monoenergetic beam of charged particles),
a scattering chamber where the beam interacts with the target, and a 
detector  for the backscattered particles. As mentioned before, the 
measured quantities are the backscattering angle $\theta$ and the energy 
$E$ of the detected particle. Good energy resolution is obviously an 
essential quantity for the accuracy of the analysis.

In the following, we give a brief description of the method.
More detailed information can be found, 
e.g., in the book by Chu, Mayer and Nicolet \cite{Chu78}.

The most important quantity determined in RBS is the kinematic 
scattering factor $k$, defined by the ratio of the energy of the 
backscattered particle $E$ and the incident energy of the projectile 
$E_0$:
\begin{equation}
k = \frac{E}{E_0} = \left( \frac{\sqrt{M_2^2-M_1^2 \sin^2\theta} 
+ M_1 \cos\theta}{M_1 + M_2} \right)^2.
\end{equation}
Here, $M_1$ and $M_2$ are the masses of the projectile and the target, 
respectively. The knowledge of the mass and energy of the projectile 
and the measurement of the energy $E$ and the angle $\theta$ of the 
backscattered particle allows the identification of the elementary 
constituents of the sample.


The thickness $t$ of the sample can be derived from the energy
loss $dE/dx$, i.e. by determining the energy of the backscattered
particles $E_1$ and $E_2$ at both edges of the sample \cite{Chu78}.

\begin{figure}[htb]
\centerline{\psfig{file=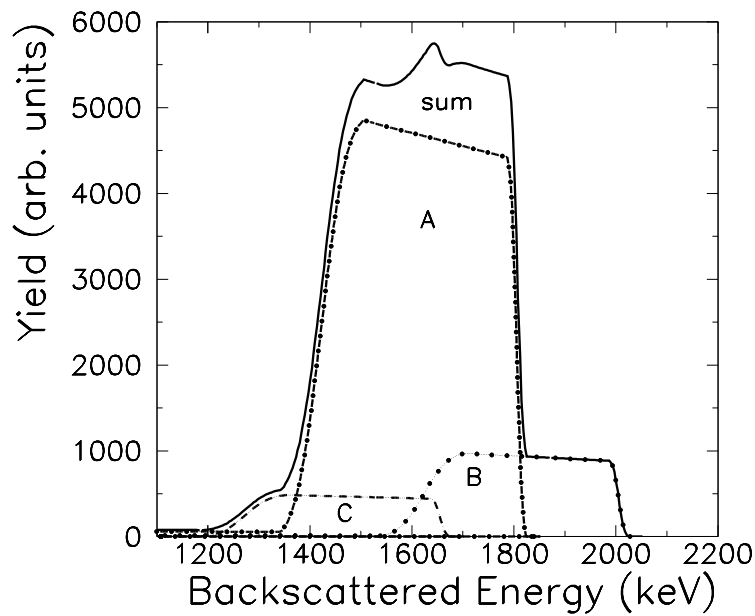,height=8.5cm}}
\caption{Schematic RBS spectrum of a sample 
which contains three different constituents (A, B, and C).
The individual contributions are shown as a dashed-dotted
(A), a dotted (B), and a dashed line (C). The sum spectrum
is displayed with a full line.
}
\label{f2}
\end{figure}

Due to the specific energy loss in different materials,
contaminations in the sample show up as distortions of 
the RBS spectrum. This is schematically shown in Fig.~\ref{f2}
for a sample which contains three different 
constituents. 
Since the amount of backscattered particles from any 
given element is proportional to its concentration,
RBS can be used to investigate quantitatively the depth
profile of individual elements in the sample.

We note that due to the strong $Z$ dependence of 
the scattering cross section, the 
RBS technique shows a lack of sensitivity for low $Z$ contaminants
imbedded in high Z materials.
RBS spectrometers using heavy ions as projectiles have been designed 
and utilized to improve the sensitivity to low $Z$ constituents 
\cite{Yu84}.

The advantages of the RBS method are many. It provides precise 
information about the sample without employing physical or chemical 
sectioning techniques and gives a quantitative analysis without references 
or standards. Furthermore, this technique is fast and non destructive.

\section{Target thickness and homogeneity}

We have utilized an RBS spectrometer at Lawrence Berkeley National
Laboratory using monoenergetic $^4$He$^+$  particles of $E_0$ = 
1.95 MeV generated by a 2.5 MeV van der Graaf accelerator. 
The diameter of the beam size was 0.75~mm.
A silicon surface barrier detector was positioned at 165$^{\circ}$
with respect to the ion beam
to collect and analyze the scattered helium particles
\cite{Yu96}.

Four different targets made of natural and isotopic lead
($^{{\rm nat},206,207,208}$Pb) have been investigated. The 
free standing targets were mounted on a thin aluminum target frame 
with a circular opening of 19~mm. The target thicknesses were 
$\sim$0.5 mg/cm$^2$. The commercially made targets were
manufactured using an evaporation method \cite{Mic}.

\begin{figure}[htb]
\centerline{\psfig{file=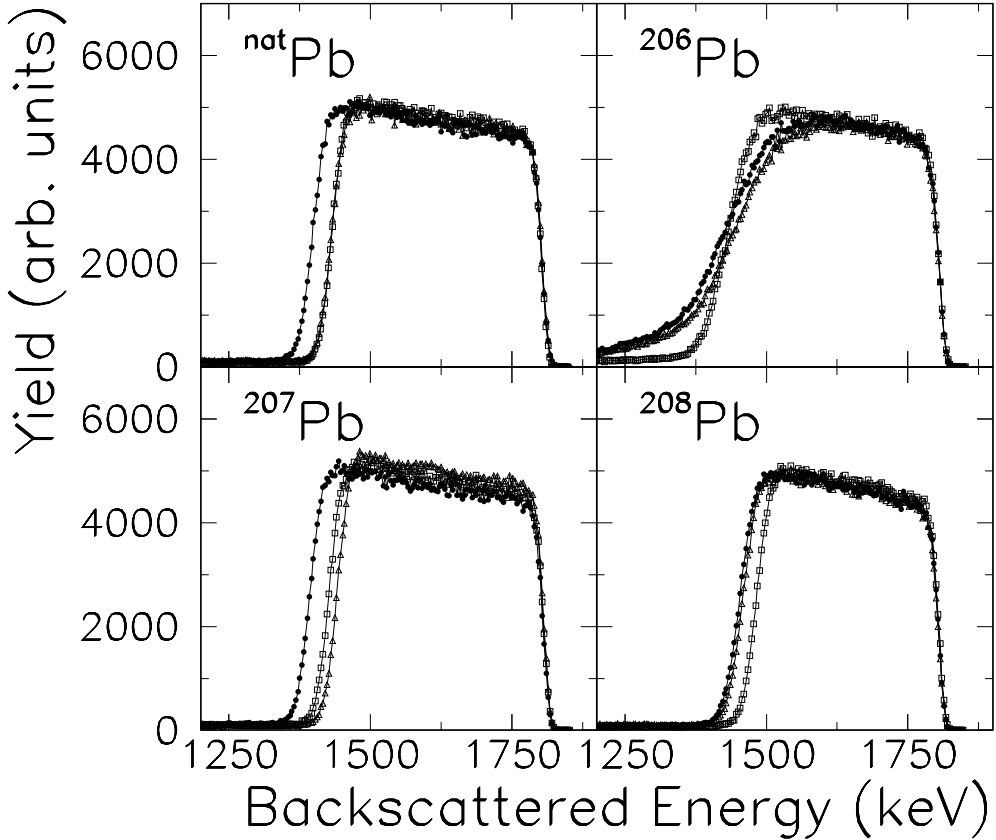,height=12cm}}
\caption{RBS energy spectra for four lead targets 
 ($^{\rm{nat}}$Pb, $^{206}$Pb, $^{207}$Pb, $^{208}$Pb).
 The different symbols correspond to different positions 
 on the target: center (full circles), upper (open squares) 
 and lower edge (open triangles).}
\label{f3}
\end{figure}

In Fig.~\ref{f3}, we show the measured energy spectra from the RBS 
analysis for four lead targets. 
The thicknesses of the foils are deduced from the
widths of the RBS spectra using the energy loss data of the
ions in Pb. The high energy edge reflects the
front and the low energy edge the back of the sample.
Small inhomogeneities in the target thickness can clearly be seen
in the figure.
In general, the spectral edges are sharply defined indicating
well defined surfaces.
In Table \ref{t1}, we compare the thicknesses determined
by direct weighing, using a geometric correction factor
to account for evaporation nonuniformity \cite{Mic},
with thicknesses determined by the RBS method.
Note that the thickness measured by RBS is given in areal density 
(atoms/cm$^2$), i.e. the amount of materials present to scatter the 
incident He ions.  This areal density can be directly compared to the 
data obtained by the direct weighing method using Avogadro's number and
the known isotopic weight of the Pb isotope.
To determine the overall homogeneity of the target,
we have measured the thickness at 3 different points
(center, lower left and upper right edge). The distance 
between the different points was 6~mm.
The standard weighing technique provides only an 
average thickness and does not provide any information  
on the homogeneity of the target foils. 
We have also calculated an average thickness using
the results from the RBS measurements according to
$ <t^{RBS}> = (t^{RBS}_{center} + t^{RBS}_{low} + t^{RBS}_{up})/3$. 
The observed agreement between the average thicknesses 
determined by the two methods is good.

\begin{figure}[htb]
\centerline{\psfig{file=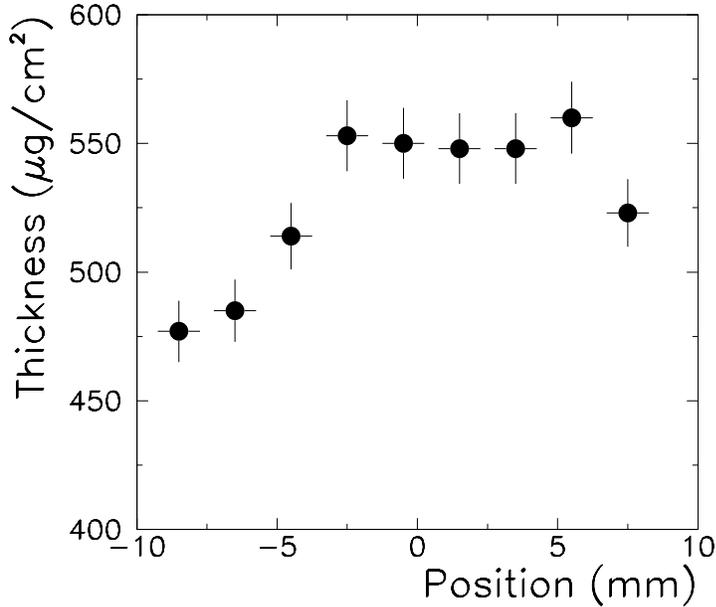,height=8.5cm}}
\caption{Target thickness determined by RBS as a function
 of position on the target surface for $^{207}$Pb.
 The error bars show the absolute uncertainty of 
 the measurement.
 }
\label{f4}
\end{figure}

In Fig.~\ref{f4}, we show the thickness as a function 
of the distance from the center 
on the surface for the $^{207}$Pb target. 
Measurements were made in 2~mm steps to determine the homogeneity.
Within the central 8-10~mm, the thickness fluctuation
is small. However, the sides are not symmetric. 
A systematic decrease of the target thickness
from the center to the edges is found which is due to
the evaporation process used to produce the target.
In our fission experiments, the diameter of the beam spot 
on the target was less than 5~mm and the accuracy of 
the center focus was $\sim$1~mm. 
Therefore, the differences in the target thickness given in Table \ref{t1} 
represent an upper limit for the uncertainty in the 
homogeneity.

\begin{table}[tb]
\caption{Target thicknesses $t$ determined from weighing in
 comparison to the results of the RBS technique. The thickness
 has been measured at three different points on the target
 (center, upper edge, lower edge). Furthermore, an average 
 thickness $<t^{RBS}>$has been calculated from these values.}
\begin{tabular}{lccccc}
 \hline
 \hline
 Target & 
 $t^{weighing}$ &
 $<t^{RBS}>$ & 
 $t^{RBS}_{center}$ & 
 $t^{RBS}_{low}$ & 
 $t^{RBS}_{up}$ \\
 &
 ($\mu$g/cm$^{2})$ &
 ($\mu$g/cm$^{2})$ &
 ($\mu$g/cm$^{2})$ &
 ($\mu$g/cm$^{2})$ &
 ($\mu$g/cm$^{2})$ \\
 \hline
  $^{nat}$Pb & 544 & 553 & 582 & 538 & 538 \\
  $^{206}$Pb & 555 & 543 & 558 & 531 & 541 \\
  $^{207}$Pb & 560 & 548 & 550 & 534 & 561 \\
  $^{208}$Pb & 500 & 490 & 503 & 496 & 472 \\
 \hline
 \hline
\end{tabular}
\label{t1}
\end{table}

We note that the relative uncertainty of the RBS target
thickness measurement is below  1\% and thus  
provides the necessary accuracy to minimize 
systematic errors in cross section measurements and associated quantities 
like, in our experiment, the first chance fission probability.

\begin{figure}[htb]
\centerline{\psfig{file=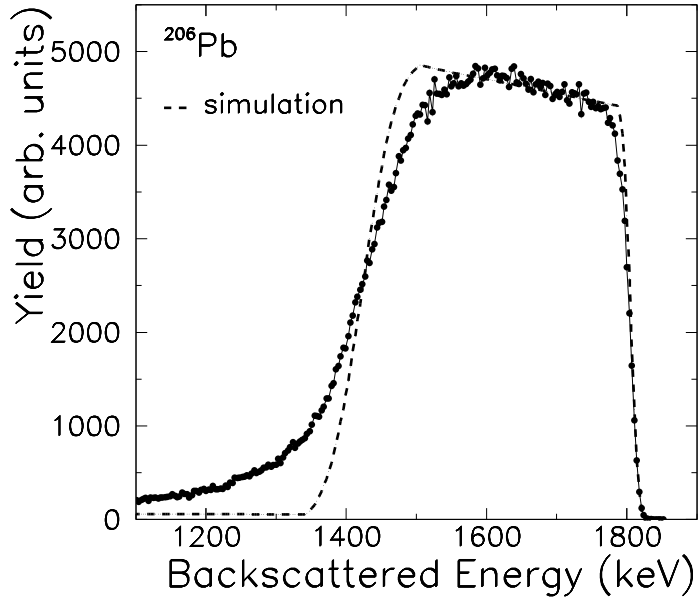,height=8.5cm}}
\caption{RBS energy spectra for $^{206}$Pb (center of target).
 The dashed line shows the results of the simulated
 spectrum. The deviations indicate surface contaminations
 and/or surface inhomogeneities.
 }
\label{f5}
\end{figure}

The sharpness of the low energy edges of the RBS energy spectra shown in
Fig.~\ref{f3} provides information on the 
surface condition, surface contaminations, and foil roughness. 
Non uniformities of the surface or significant contaminations
will broaden the energy of the backscattered $^4$He particle.
For all targets the front edges are very sharp and for
three targets the back edges are also quite sharp. 
However, for the 
$^{206}$Pb case, the back edge is significantly 
washed out compared to other targets. 
In Fig.~\ref{f5}, we show $^{206}$Pb data in comparison to
the simulated sum spectrum obtained from the RBS 
analysis \cite{Saa92}.
While the agreement is good for the front edge and 
inside the target material,
a large deviation is observed at the back edge.
This is most likely caused by either 
a significant surface inhomogeneity or the presence of 
large particles in the foil.

\section{Target impurities}

In order to determine whether any significant
target impurities were present, particle induced x-ray 
emission (PIXE) has been measured simultaneously during
the RBS experiments. 
PIXE is an analytical method which relies on the spectrometry 
of characteristic x-rays emitted by the target atoms due to
the irradiation with a high energy ion beam. The method can
identify various constituents in a compound target
via their characteristic x-rays.
To measure the x-rays,  we have used a lithium drifted silicon
detector which was located at 30$^{\circ}$ with respect to the
incident beam. Under the most favorable conditions, a detection limit 
of $\sim$ 1~ppm  for thin foils can be achieved \cite{Yu96}.
Compared to RBS, this method is significantly more sensitive
to determine target impurities \cite{Yu96}.

\begin{figure}[htb]
\centerline{\psfig{file=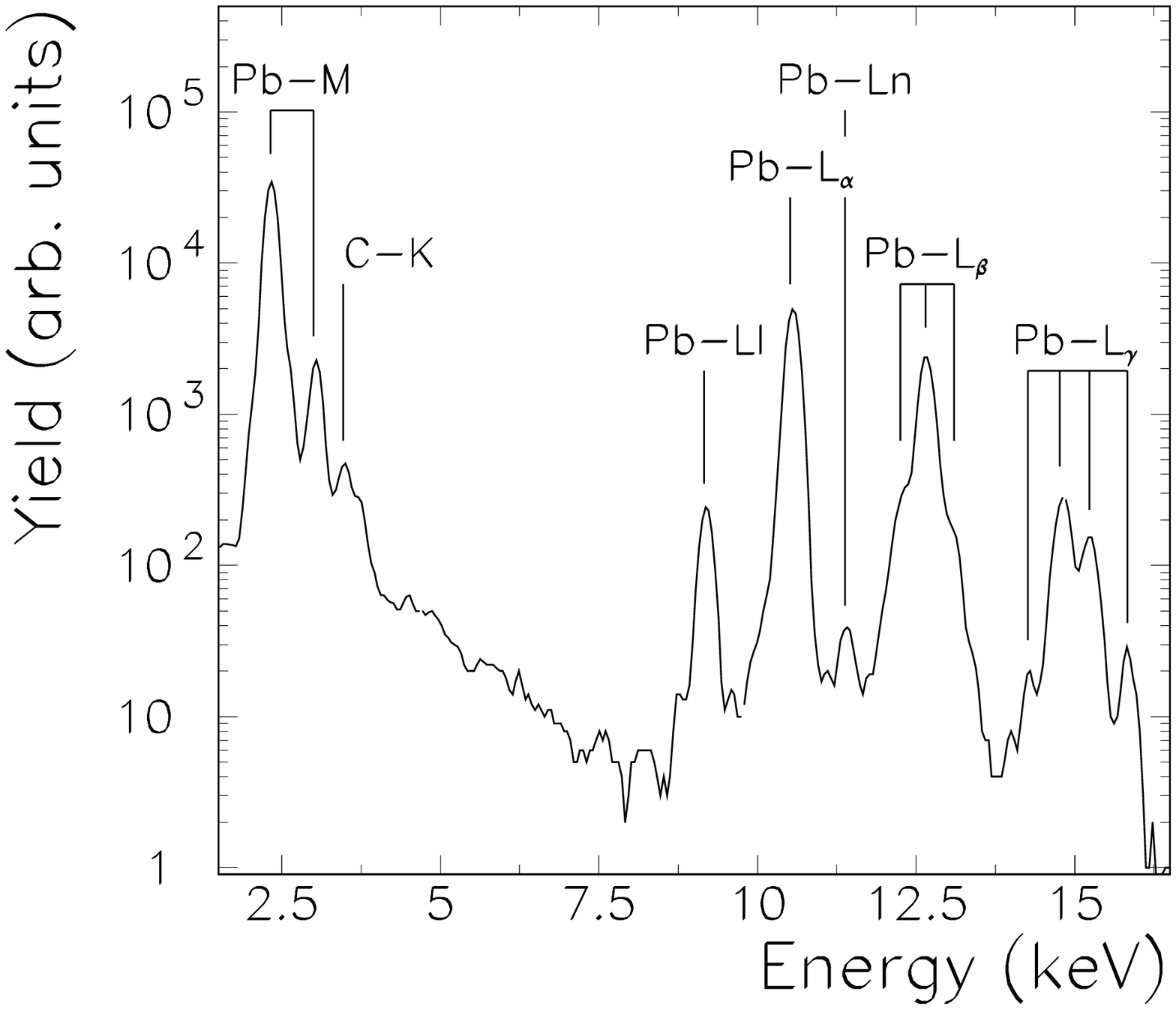,height=8.5cm}}
\caption{Particle induced x-ray emission (PIXE) spectrum
 for $^{207}$Pb (center of target).
 }
\label{f6}
\end{figure}

In Fig.~\ref{f6}, we show the accumulated x-ray spectrum for one
of the targets ($^{207}$Pb). The spectrum is dominated by
the various M and L x-ray peaks of Pb confirming that
Pb is the major constituent.
In addition, a small peak from the carbon backing of the target
is seen. No sizable contribution of other contaminations  has 
been detected. We note that for the present experimental 
conditions the detectable limit for most transition metals
is 10-50~ppm.


\section{Relative cross sections}

A good relative accuracy of the measured fission cross sections of
the neighboring compound nuclei is very important to minimize
the associated error in measurements of first chance fission cross 
sections. To check this quantity for several separated isotopic targets, 
we have applied an independent method based on the measurement 
of the cross section of the corresponding natural target. 

In our experiment, we have measured the fission cross sections
of four different lead targets ($^{206,207,208}$Pb and $^{\rm nat}$Pb).
The composition of natural lead is: 52.4\% of $^{208}$Pb,
22.1\% of $^{207}$Pb, 24.1\% of $^{206}$Pb, and 1.4\% of $^{204}$Pb.
Unfortunately, we have not measured the fission cross section
of the latter isotope and had to estimate it from the
ratio of the cross sections for $^{208}$Pb and $^{206}$Pb. 
This estimate is in agreement with measured fission cross sections
for all three isotopes \cite{Kho66}.
We have calculated the ``natural'' cross section by 
adding up the relative isotopic cross sections using the target
thicknesses determined by RBS:
\begin{equation}
 \sigma_{\rm nat}^{\rm calc} = \sum_{i=204}^{208} p_{i} \sigma_{i}.
\label{nat_xs}
\end{equation}
Here, $p$ represents the contribution of the isotope $i$ to
the natural composition. 

\begin{figure}[htb]
\centerline{\psfig{file=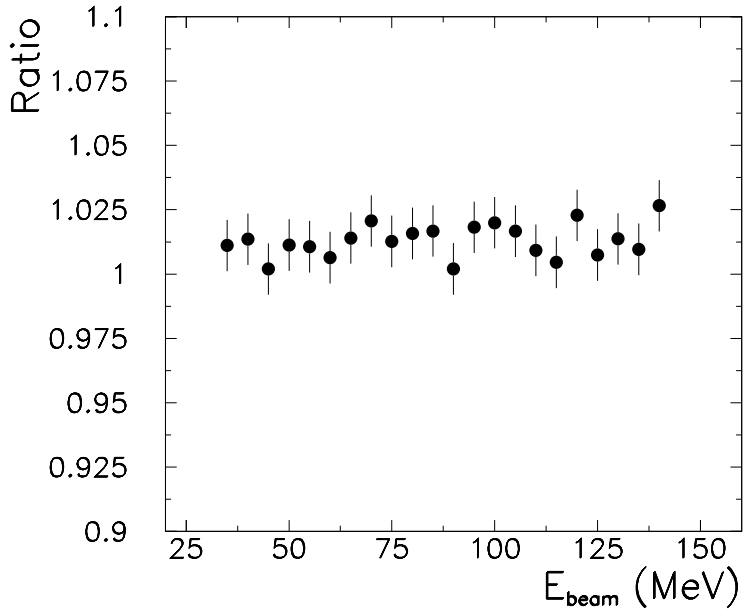,height=8.5cm}}
\caption{Ratio of the calculated fission cross section
 for natural lead using the individually measured 
 cross sections of the lead isotopes and the measured cross section
 using a natural lead target. The projectile is $^3$He.
}
\label{f7}
\end{figure}

In Fig.~\ref{f7}, we show the results of this analysis; the  
calculated cross sections from Eq.~\ref{nat_xs} have been 
normalized by the cross section measured for the natural lead
target. 
A rather constant value close to unity has been found.
This good agreement
allows us to conclude that the relative cross sections 
are known to $\pm$2\%. 
This accuracy is a substantial improvement over 
previous experiments \cite{Kho66} and is sufficiently
good to allow extraction of first chance fission 
probabilities from our data \cite{Rub96c}.

\section{Summary}

In this paper, we have presented results of a 
method that allows precise characterization
of thin target foils used in nuclear physics experiments. 
The applied Rutherford backscattering and particle induced 
x-ray emission techniques provide
information on the thickness, homogeneity,
and constituents of a target
material. Furthermore, this method is fast and -- more
importantly -- non-destructive.

The information allows one to minimize systematic errors due to 
uncertainties in the target thickness and homogeneity. 
The technique described in this paper thus 
provides a powerful tool to determine 
the purity of a target and is especially useful if it is applied 
in advance of an experiment.

\bigskip    
This work was supported by the Director, Office of Energy Research,
Office of High Energy and Nuclear Physics, Nuclear Physics Division
of the US Department of Energy, under contract DE-AC03-76SF00098.

\end{document}